\documentclass[fleqn,10pt]{wlscirep}
\usepackage[utf8]{inputenc}
\usepackage[T1]{fontenc}

\newcommand{\Rm}{\mathbb{R}}

\newcommand{\be}{\[}
\newcommand{\ee}{\]}
\newcommand{\ba}{\[\begin{aligned}}
\newcommand{\ea}{\end{aligned}\]}

\newcommand{\bv}[1]{\boldsymbol{\mathrm{#1}}}

\title{Prediction of cerebral blood volume change after resuscitation from hypoxic-ischemic insult for newborn piglets}

\author[1,*]{Manabu Machida}
\author[2,3]{Tsutomu Mitsuie}
\author[4]{Shinji Nakamura}
\author[4]{Takashi Kusaka}
\affil[1]{Department of Informatics, Faculty of Engineering, Kindai University, Higashi-Hiroshima 739-2116, Japan}
\affil[2]{Medical Engineering Equipment Management Center, Kagawa University Hospital, Kagawa, Japan}
\affil[3]{Graduate School of Medicine, Faculty of Medicine, Kagawa University, Kagawa, Japan}
\affil[4]{Department of Pediatrics, Faculty of Medicine, Kagawa University, Kagawa, Japan}

\affil[*]{machida@hiro.kindai.ac.jp}

\begin{abstract}
Neonatal hypoxic-ischemic encephalopathy (HIE) is a significant cause of neonatal mortality and developmental disabilities. It has been revealed that the temporal behavior of the cerebral blood volume (CBV) carries information on the degree of hypoxia-ischemia. CBV can be estimated by means of near-infrared spectroscopy. The change of CBV after the insult is related to the change of CBV during the insult. In this paper, we consider a mathematical model which governs the time evolution of CBV after the insult. We show that the temporal behavior of CBV can be predicted with the Kalman filter which is based on the mathematical model.
\end{abstract}
\begin{document}

\flushbottom
\maketitle
%
%
\thispagestyle{empty}

\section*{Introduction}
\label{intro}

Neonatal hypoxic-ischemic encephalopathy (HIE) causes of death and developmental disabilities in newborns. Therapeutic hypothermia is often not effective. Thus it is necessary to understand the degree of hypoxia-ischemia (HI) to make treatment plans \cite{Htun-etal19,Bonifacio-etal15,Gunn-Bennet09}. Treatments must be delivered within six hours of birth. Hence early recognition of cerebral hemodynamic changes is necessary.

Since near-infrared light is particularly absorbed by hemoglobin, cerebral blood volume (CBV) can be estimated with near-infrared spectroscopy (NIRS) using the time-resolved spectroscopy (TRS, Hamamatsu Photonics K.K.). For the piglet experiment which will be considered in this paper, two optical fibers (one is for emission and the other is for detection) were attached to the head of each piglet with the source-detector distance $30\,{\rm mm}$. In TRS, the time-correlated single-photon counting technique is used to detect photons. By photon detection with three wavelengths, the oxyHb and deoxyHb concentrations can be estimated, which can then be converted to CBV \cite{Ijichi-etal05a,Ijichi-etal05b}. Compared with X-ray, light is free from radiation exposure. This means that NIRS measurements can be performed repeatedly. Thus a time series of NIRS imaging is available. The use of NIRS for monitoring oxygenation of the brain has provided useful insights for the management of newborns \cite{Dix-etal17,Kusaka-etal14}.

An asphyxia piglet model has been established, which exhibits a uniform degree of histopathological brain injury, and the change in CBV was investigated for the piglets by means of near-infrared imaging \cite{Nakamura-etal13,Nakamura-etal14,NakamuraM-etal15,Nakamura-etal15}. In piglets subjected to the HI insult, CBV initially rises to a peak and then keeps decreasing until resuscitation. The CBV reduction from the peak during the insult is related to the severity of brain injuries caused by autoregulatory impairment. Impaired cerebral autoregulation results in adverse neurological outcomes \cite{Ferriero04,Alderliesten-etal13}. The time evolution of CBV reflects the degree of hypoxia-ischemia. In particular, the dependence of the temporal change $y$ after the insult on the decay amount $x$ of CBV during the HI insult was investigated \cite{Mitsuie-etal21}.

To predict the time evolution of CBV after the insult, it is desirable to find a differential equation which governs the phenomenon. The change of CBV in time takes place as a consequence of extremely complicated blood flows in the body, which in principle can be described by the fluid dynamics. Without touching detailed blood flows in the head, we seek a differential equation which is able to reproduce the time evolution of CBV. This is in some sense related to seeking laws of thermodynamics without statistical mechanics. In this paper, we propose a differential equation which governs the time evolution of CBV for the piglets. To validate the mathematical model, we predict the time dependence of CBV using the observed data from the piglet experiments. The Kalman filter was used for the prediction.

The Kalman filter, including the extended Kalman filter, has been used for a pulmonary blood flow estimator \cite{Brovko81}, cerebral blood flow autoregulation \cite{Masnadi-Shirazi-etal04,Aoi-etal09}, and the heart rate detection with reflected light \cite{Prakash-Tucker18}, microvessel imaging for microvessel density maps and blood flow speed maps \cite{Tang-etal20}, and the hemodynamic responses for functional near-infrared spectroscopy \cite{Dong-etal19}. See a review \cite{Nolte-Bertoglio22} for the blood flow for the cardiovascular system.

Since it is not easy to obtain the time dependence of CBV for neonates and even newborn piglets, the Kalman filter has not been used in the context of neonatal HIE. Moreover, to the best of our knowledge, any mathematical model has not been proposed for the time evolution of CBV for the neonatal HIE.

\section*{Results}

\begin{figure}[ht]
\centering
\includegraphics[width=0.45\linewidth]{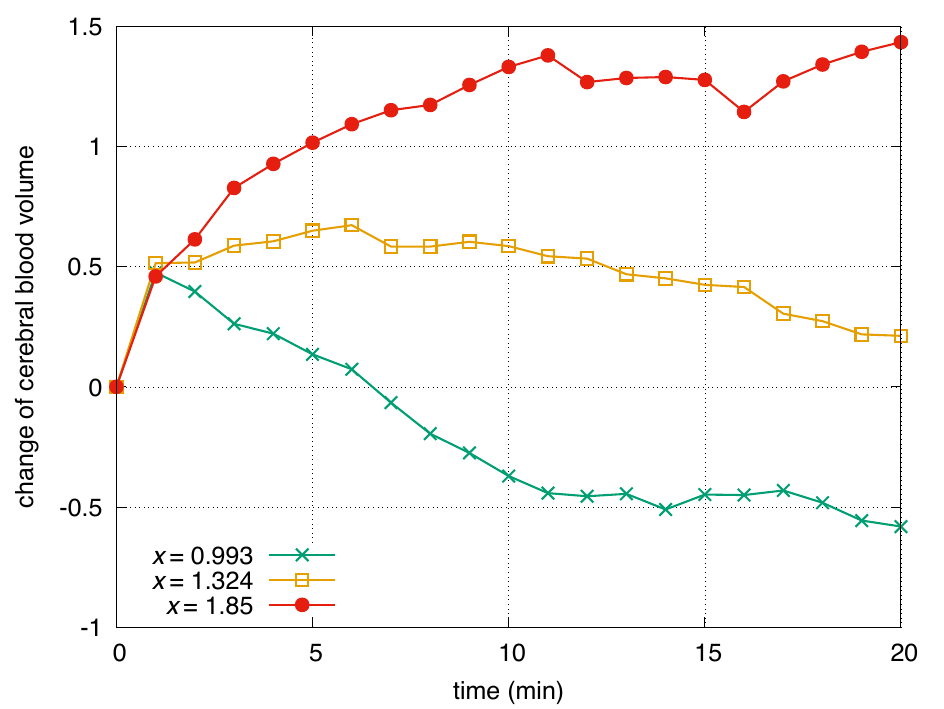}
\caption{Time evolutions of the change $y$ of CBV for three cases which are used in this paper.}
\label{fig1}
\end{figure}

Figure \ref{fig1} shows the temporal change $y$ of CBV beginning at the resuscitation from the HI insult for different $x$. See Ref.~\citenum{Mitsuie-etal21} for piglet experiments for the data shown in Fig.~\ref{fig1}. Measurements were performed every $10\,{\rm sec}$ and averages of six measurements were taken to plot the temporal profile every minute. In Fig.~\ref{fig1}, the red line with solid circles is $y$ for $x=1.85$, the ocher line with open squares is $y$ for $x=1.324$, and the green line with diagonal crosses is $y$ for $x=0.993$.

\begin{figure}[ht]
\centering
\includegraphics[width=0.45\linewidth]{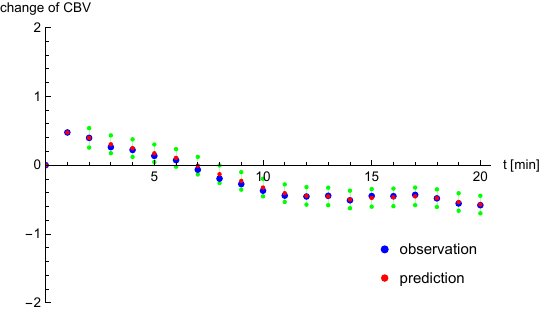}
\hspace{5mm}
\includegraphics[width=0.45\linewidth]{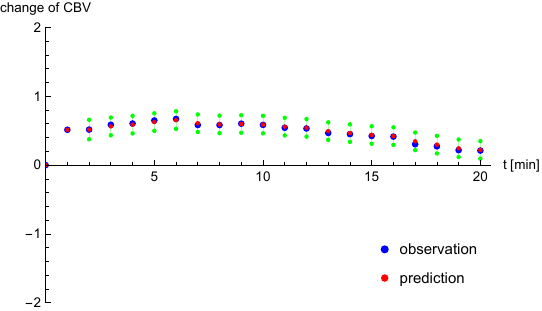}
\\ \vspace{5mm}
\includegraphics[width=0.45\linewidth]{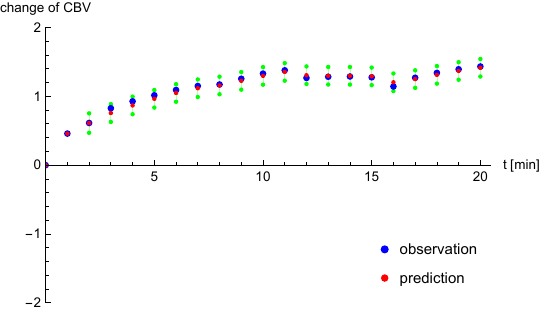}
\caption{One-step ($1\,{\rm min}$) predictions for (top left) $x=0.993$, $\omega=0.446$, (top right) $x=1.324$, $\omega=0.370$, and (bottom) $x=1.85$, $\omega=0.243$. Blue circles show the observed data and red circles mean predicted values. Moreover, green circles show the standard deviation of the estimated error in upper and lower directions.}
\label{fig2}
\end{figure}

For each curve in Fig.~\ref{fig1}, one-step prediction was done with the Kalman filter. Curves in Fig.~\ref{fig1} are drawn with $N+1$ points $(t_i,y_i)$ ($i=0,1,\dots,N$). Here, $N=20$. We note that $t_i-t_{i-1}=1\,{\rm min}$ ($i=1,\dots,N$). In each of five panels of Fig.~\ref{fig2}, predicted values of $y_i$ were computed with the Kalman filter using $y_1,\dots,y_{i-1}$. The linear mathematical model (see below) was used for the Kalman filter. In this way, the actual values (red) and predicted values (blue) are compared in each panel. In the panels of Fig.~\ref{fig2}, green points show the standard deviation of the error at each time. In the case of $x=0.993$, $\omega=0.446\,{\rm rad/min}$. In the case of $x=1.324$, $\omega=0.370\,{\rm rad/min}$. In the case of $x=1.85$, $\omega=0.243\,{\rm rad/min}$. In all cases, $t_c=1\,{\rm min}$, $x_p=1.5$, $a=1$, $b=0.3$. The unit of time was hour instead of minute for the Kalman filter, in which case $\Delta\tau=1/60\,{\rm hr}$.

\begin{figure}[ht]
\centering
\includegraphics[width=0.45\linewidth]{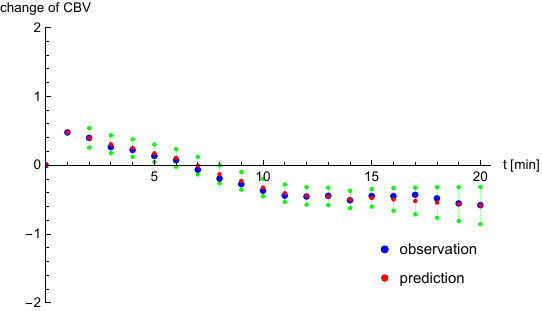}
\hspace{5mm}
\includegraphics[width=0.45\linewidth]{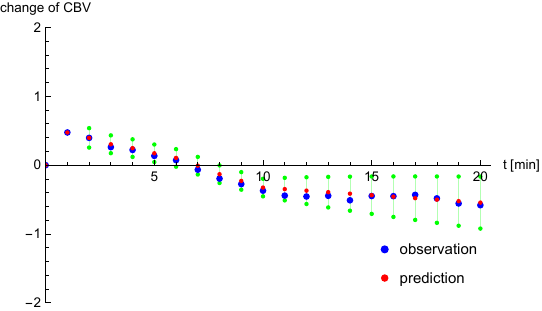}
\\
\vspace{5mm}
\includegraphics[width=0.45\linewidth]{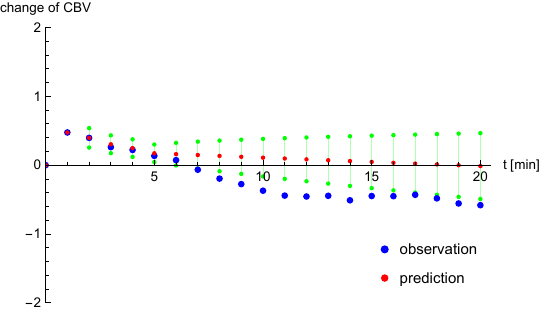}
\caption{Predictions for the last (top left) $5\,{\rm min}$, (top right) $10\,{\rm  min}$, and (bottom) $15\,{\rm  min}$ in the case of $x=0.993$, $\omega=0.446$.}
\label{fig3}
\end{figure}

\begin{figure}[ht]
\centering
\includegraphics[width=0.45\linewidth]{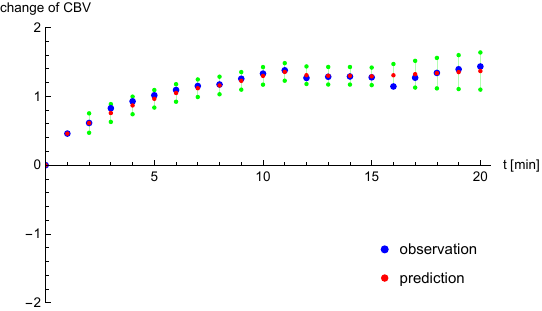}
\hspace{5mm}
\includegraphics[width=0.45\linewidth]{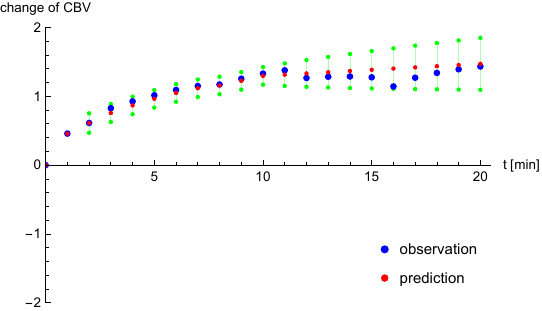}
\\
\vspace{5mm}
\includegraphics[width=0.45\linewidth]{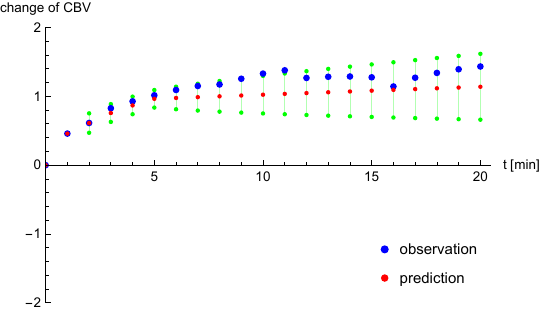}
\caption{Predictions for the last (top left) $5\,{\rm min}$, (top right) $10\,{\rm  min}$, and (bottom) $15\,{\rm min}$ in the case of $x=1.85$, $\omega=0.243$.}
\label{fig4}
\end{figure}

In Figs.~\ref{fig3} and \ref{fig4}, prediction for longer times was considered. Figure \ref{fig3} shows an example of $y$ which decays in time and Fig.~\ref{fig4} shows an example of $y$ which grows in time. In the top left panels of Figs.~\ref{fig3} and \ref{fig4}, predicted values are plotted for the last $5\,{\rm min}$. That is, the Kalman filter algorithm did not use observed data for the last $5\,{\rm min}$. Similarly, in the top right panels of Figs.~\ref{fig3} and \ref{fig4}, predicted values for the last $10\,{\rm min}$ are plotted. In the bottom panels of Figs.~\ref{fig3} and \ref{fig4}, predicted values for the last $15\,{\rm min}$ are plotted.

\begin{figure}[ht]
\centering
\includegraphics[width=0.45\linewidth]{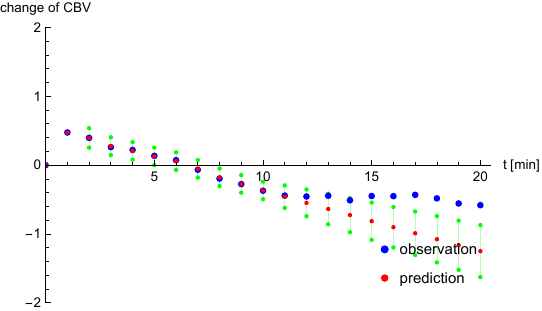}
\hspace{5mm}
\includegraphics[width=0.45\linewidth]{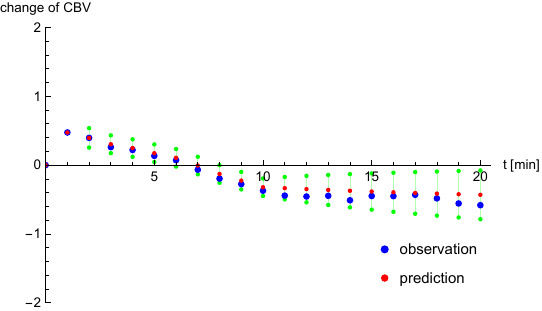}
\caption{These graphs correspond to the top right panel of Fig.~\ref{fig3}. In the case of $x=0.993$, (left) $a=10$ and (right) $b=3$ while other parameters are the same as the parameters used in Fig.~\ref{fig3}.
}
\label{fig5}
\end{figure}

The parameters in the proposed mathematical model were chosen to reproduce the observed behavior of $y$. In particular, the same $\beta,b$ were used for Figs.~\ref{fig2}, \ref{fig3}, and \ref{fig4}. Figure \ref{fig5} illustrates predicted values when other $\beta,b$ are used.

\begin{figure}[ht]
\centering
\includegraphics[width=0.3\linewidth]{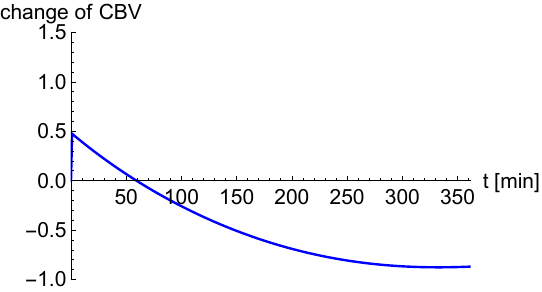}
\hspace{2mm}
\includegraphics[width=0.3\linewidth]{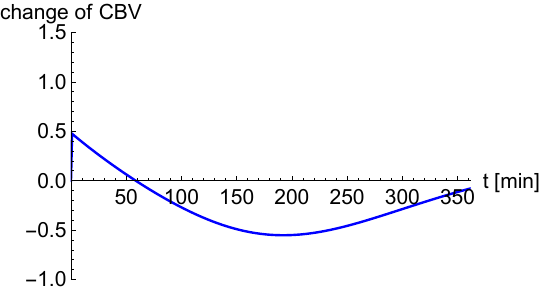}
\hspace{2mm}
\includegraphics[width=0.3\linewidth]{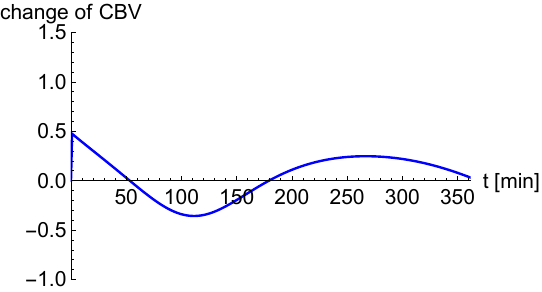}
\\
\vspace{5mm}
\includegraphics[width=0.3\linewidth]{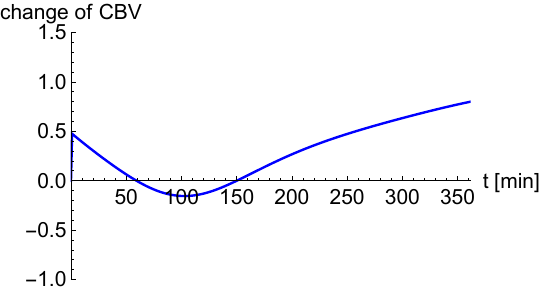}
\hspace{2mm}
\includegraphics[width=0.3\linewidth]{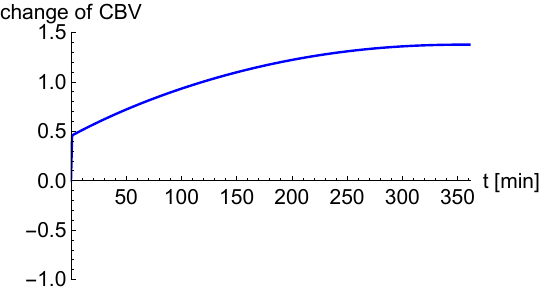}
\hspace{2mm}
\includegraphics[width=0.3\linewidth]{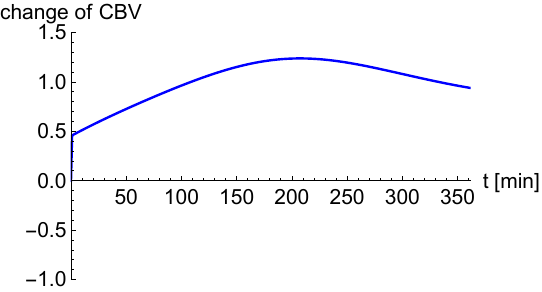}
\caption{
(upper left) $x=0.993$, $\gamma=1.0$, $\beta=0.01$, 
(upper center) $x=0.993$, $\gamma=1.0$, $\beta=0.1$, 
(upper right) $x=0.993$, $\gamma=1.0$, $\beta=1.0$, 
(lower left) $x=0.993$, $\gamma=0.5$, $\beta=1.0$, 
(lower center) $x=1.85$, $\gamma=1.5$, $\beta=0.01$,
(lower right) $x=1.85$, $\gamma=1.5$, $\beta=0.1$. 
For all cases, $b=0.3$ and $t_c=1\,{\rm min}$.
}
\label{fig6}
\end{figure}

The nonlinear mathematical model was used for panels in Fig.~\ref{fig6}: 
(upper left) $x=0.993$, $b=0.3$, $\gamma=1.0$, $\beta=0.01$, 
(upper center) $x=0.993$, $b=0.3$, $\gamma=1.0$, $\beta=0.1$, 
(upper right) $x=0.993$, $b=0.3$, $\gamma=1.0$, $\beta=1.0$, 
(lower left) $x=0.993$, $b=0.3$, $\gamma=0.5$, $\beta=1.0$, 
(lower center) $x=1.85$, $b=0.3$, $\gamma=1.5$, $\beta=0.01$,
(lower right) $x=1.85$, $b=0.3$, $\gamma=1.5$, $\beta=0.1$. 
For all cases, $t_c=1\,{\rm min}$.

\section*{Discussion}

As shown in Fig.~\ref{fig1}, the change $y$ of CBV has an increasing tendency, a decreasing tendency, or no strong increasing nor decreasing behavior. The major cause of such variety is the amount $x$ of the drop of CBV during the HI insult between the peak time and the resuscitation time \cite{Mitsuie-etal21}. During the HI insult, CBV initially rises rapidly as a compensatory response, followed by impaired cerebral blood flow autoregulation and vasoparalysis, leading to a gradual decrease in CBV as decompensation sets in. It was found that a large decrease of CBV from the baseline during the insult was associated with severe brain injury or mortality \cite{Mitsuie-etal21}.

Since the change of $y$ during $1\,{\rm min}$ is monotonic, the one-step prediction (prediction every minute) worked well as shown in Fig.~\ref{fig2}. In Fig.~\ref{fig2}, predicted values agree with the observed data after the first a few minutes of the filtering process of the Kalman filter.

Longer predictions were performed in Figs.~\ref{fig3} and \ref{fig4}. We can observe that predictions within $10\,{\rm min}$ work quite well. Since longer predictions were tested using the same time-series data up to $20\,{\rm min}$, the filter process becomes shorter for longer prediction. Moreover, since the linear model is used for the Kalman filter, it is not possible to predict nonmonotonic behavior of CBV.

We properly chose parameters in the mathematical model. This is explained in Fig.~\ref{fig5}. The prediction is not successful if other parameters are chosen. We emphasize that the parameter $x,\omega$ depend on individual sample but otherwise the same parameters  were used for all samples in Figs. \ref{fig2} through \ref{fig4}.

In the prediction of CBV, nonmonotonic behavior of $u$ can be obtained if we use the nonlinear mathematical model, which has a nonlinear term (see below). Examples of calculated $u$ are shown in Fig.~\ref{fig6}. As shown in Fig.~\ref{fig1}, the change $y$ has a decreasing tendency for the first $20\,{\rm min}$ for $x=0.993$ and has a increasing tendency for $x=1.85$. Their behaviors might not be monotonic for a long time of a few or several hours. The nonlinear model can produce different time-evolution patterns as shown in Fig.~\ref{fig6} depending of the choice of $\gamma$, $\beta$.

In principle, parameters in the mathematical models depend on factors such as blood pressure and heart rate, as well as sex differences, presence or absence of infections, and severity. For example, it was observed for both piglets \cite{Mitsuie-etal21} and fetal sheep \cite{Yamaguchi-etal18} that if the heart rate increases more during the insult, it also increases more after the insult.

\section*{Conclusion}

In this paper, we explored the change of CBV during $20$ minutes after the HI insult in neonatal piglets with HIE. In general, a large CBV decrease during the insult causes an increase of the post-insult CBV.

We proposed a mathematical model which can reproduce the behavior that greater decreases in CBV during the HI insult are followed by more pronounced increases after the insult.

Based on the proposed linear model, we showed that the time evolution of CBV can be predicted with the Kalman filter. This fact implies that the proposed model governs the change of CBV in time for neonatal HIE.

\section*{Materials and methods}

\subsection*{Model}

Let $V(t)$ be the cerebral blood volume (${\rm mL}\,/\,100\,{\rm g}$) at time $t\,({\rm hr})$. We suppose that the hypoxic-ischemic period ends at $t=0$. Let $t_{\rm max}>0$ be the maximum time ($t_{\rm max}=1/3$ in this paper). We can express $V(t)$ as
\be
V(t)=v_{\rm max}-x+y(t;x),\quad 0<t<t_{\rm max},
\ee
where $x$ is the blood volume which goes away from the brain after it reaches the peak $v_{\rm max}$ until $t=0$ and $y(t;x)$ is the change of the cerebral blood volume for $t>0$.

In light of the above discussion, we make the following new model:
\be
y(t;x)=\left\{\begin{aligned}
x\tan(\omega t),&\quad 0<t\le t_c,
\\
u(\tau;x)+y(t_c;x),&\quad \tau=t-t_c,\quad t_c<t<t_{\rm max},
\end{aligned}\right.
\ee
where $\omega$ is a positive constant. We give $\xi(x)$ as
\be
\xi(x)=\frac{x-x_p}{\cos(\omega t_c)},
\ee
where $(x_p,y_p)$ satisfies
\be
\omega t_c=\tan^{-1}\left(\frac{y_p}{x_p}\right).
\ee

\subsubsection*{Linear model}

Let us consider the equation of motion in the presence of friction:
\be
\frac{d^2u}{d\tau^2}=-b\frac{du}{d\tau},\quad b>0.
\ee
We give the initial conditions as
\be
u(0)=0,\quad
\frac{du}{d\tau}(0)=a\xi(x),
\ee
where $a$ is a positive constant.

The solution to the linear model is obtained as
\be
u(\tau)=\frac{a}{b}\xi(x)\left(1-e^{-bt}\right).
\ee

\subsubsection*{Nonlinear model}

Nonmonotonic behavior of $u$ can be obtained if the motion is considered in a potential $W$,
\be
\frac{d^2u}{d\tau^2}=-b\frac{du}{d\tau}-\frac{dW}{du},\quad b>0,
\ee
where
\be
W=\frac{\beta}{4\xi(x)^2}u^4-\frac{\beta\gamma^2}{2}u^2,\quad
\beta>0,\quad\gamma>0.
\ee
We note that the double-well potential $W$ has two minima at $u=\pm\gamma\xi$, and one local maximum at $u=0$. From the equation of motion we obtain the following Duffing-type equation.
\be
\left\{\begin{aligned}
\frac{d^2u}{d\tau^2}+b\frac{du}{d\tau}-\beta\gamma^2u+\frac{\beta}{\xi(x)^2}u^3
&=0,\quad\tau>0,
\\
u(0)&=0,
\\
\frac{du}{d\tau}(0)&=a\xi(x).
\end{aligned}\right.
\ee
We note that $\beta\gamma^2$ means stiffness, and $\beta/\xi^2$ indicates the degree of nonlinearity.

\subsection*{The Kalman filter}

The Kalman filter is described below using the linear mathematical model. We have
\be
\left\{\begin{aligned}
\frac{d}{d\tau}\begin{pmatrix}u\\v\end{pmatrix}
&=
\begin{pmatrix}v\\-bv\end{pmatrix},\quad\tau>0,
\\
\begin{pmatrix}u(0)\\v(0)\end{pmatrix}
&=
\begin{pmatrix}0\\a\xi(x)\end{pmatrix}.
\end{aligned}\right.
\ee
Let us discretize time as $t_k=k\Delta\tau$ ($t_c=k_c\Delta\tau$). We can write
\be
\begin{pmatrix}u_k\\v_k\end{pmatrix}=
\begin{pmatrix}
u_{k-1}+v_{k-1}\Delta\tau \\ v_{k-1}-bv_{k-1}\Delta\tau
\end{pmatrix}
\ee
for $k=k_c+1,\dots$. Let us set
\be
F=\begin{pmatrix}1 & \Delta\tau \\ 0& 1-b\Delta\tau\end{pmatrix},
\quad
H=\begin{pmatrix}1 & 0\end{pmatrix}.
\ee
The state equation and observation equation are given by
\be
\bv{x}_k=F\bv{x}_{k-1}+\bv{w}^{(1)},\quad
y_k(x)=H\bv{x}_k+w^{(2)},
\ee
where $\bv{w}^{(1)}\in\Rm^2$ and $w^{(2)}\in\Rm$ are Gaussian noise for the system and measurement noise. Let $Q$, $R$ be diagonal system noise and measurement error covariance matrices:
\be
Q=E\left[\bv{w}^{(1)}\bv{w}^{(1)T}\right]\in\Rm^2,\quad
R=E\left[w^{(2)2}\right]\in\Rm,
\ee
where $E[\cdot]$ denotes the expectation value. We set $Q=0.01I$, $R=0.01$, where $I$ is the $2\times 2$ identity matrix.

Let $\hat{\bv{x}}_k$ be the estimated state vector. The calculation consists of the prediction and update steps \cite{Simon}. The prediction step is done as follows. We have
\be
\Rm^2\ni \hat{\bv{x}}_{k|k-1}=F\hat{\bv{x}}_{k-1},\quad
\Rm^{2\times 2}\ni P_{k|k-1}=FP_{k-1}F^T+Q.
\ee
Here, the error covariance matrix $P_k$ is given by
\[
P_k=E\left[(\bv{x}_k-\hat{\bv{x}}_k)(\bv{x}_k-\hat{\bv{x}}_k)^T\right].
\]
Initially we set $P_{k_c}=I$. Then the update step is performed as follows:
\ba
\Rm\ni z_k&=y_k-H\hat{\bv{x}}_{k|k-1},
\\
\Rm\ni S_k&=HP_{k|k-1}H^T+R=\{P_{k|k-1}\}_{11}+R,
\\
\Rm^2\ni K_k&=P_{k|k-1}H^TS_k^{-1}=
\begin{pmatrix}\{P_{k|k-1}\}_{11}\\\{P_{k|k-1}\}_{21}\end{pmatrix}\frac{1}{S_k},
\\
\Rm^2\ni \hat{\bv{x}}_k&=\hat{\bv{x}}_{k|k-1}+K_kz_k,
\\
\Rm^{2\times 2}\ni P_k&=\left(I-K_kH\right)P_{k|k-1}.
\ea

Let us consider the $j$-step prediction. After $k=k_M$, we have
\ba
\hat{\bv{x}}_{k_M+i}
&=
F\hat{\bv{x}}_{k_M+i-1},
\\
P_{k_M+i}&=FP_{k_M+i-1}F^T+Q
\ea
for $i=1,\dots,j$.

\section*{Author contributions}
M.M. designed the study and drafted the article. T.M., S.N., and T.K. performed the animal experiments. M.M. performed the data analysis. M.M., T.M., S.N., and T.K. critically revised the manuscript and contributed to the final approval of the version to be published.

\section*{Funding}
This work was supported by JST, PRESTO Grant Number JPMJPR2027. This study was also supported by JSPS KAKENHI Grant Numbers JP22K07822, JP23K07332.

\section*{Competing interests}

The authors declare no competing interests.

\section*{Additional information}

Correspondence and requests for materials should be addressed to M.M.

\end{document}